\documentclass{pasa}%

\usepackage{graphicx}
\usepackage[dvipsnames]{xcolor}
\usepackage{mathtools}
\usepackage{amsmath}
\usepackage{multicol,lipsum}
\usepackage{xcolor}
\title[Non-Coherent Survey of LEO using MWA]{A Low Frequency Blind Survey of the Low Earth Orbit Environment using Non-Coherent Passive Radar with the Murchison Widefield Array}

\author[Prabu et al.]{Prabu, S.$^{1,3}$, Hancock, P.$^1$, Zhang, X.$^2$ and Tingay, S.J.$^{1}$
\affil{$^1$International Centre for Radio Astronomy Research, Curtin University, Bentley, WA 6102, Australia}%
\affil{$^2$CSIRO Astronomy and Space Science, 26 Dick Perry Avenue, Kensington, WA 6151, Australia}
\affil{$^3$CSIRO Astronomy and Space Science, Corner Vimiera \& Pembroke Roads, Marsfield, NSW 2122, Australia}
}%

\jid{PASA}
\doi{10.1017/pas.\the\year.xxx}
\jyear{\the\year}
\usepackage{longtable}
\usepackage{aas_macros}
\usepackage{hyperref} 
\hypersetup{colorlinks,citecolor=blue,linkcolor=blue,urlcolor=blue}

\definecolor{customColor1}{RGB}{0, 0, 0}
\definecolor{customColor2}{RGB}{187, 116, 131}
\definecolor{customColor3}{RGB}{50, 205, 50}
\definecolor{customColor4}{RGB}{201, 124, 178}

\begin{document}

\begin{frontmatter}
\maketitle

\begin{abstract}
We have extended our previous work to use the Murchison Widefield Array (MWA) as a non-coherent passive radar system in the FM frequency band, using terrestrial FM transmitters to illuminate objects in Low Earth Orbit LEO) and the MWA as the sensitive receiving element for the radar return. We have implemented a blind detection algorithm that searches for these reflected signals in difference images constructed using standard interferometric imaging techniques.  From 20 hours of archived MWA observations, we conduct a survey of LEO, detecting 74 unique objects over multiple passes and demonstrating the MWA to be a valuable addition to the global Space Domain Awareness network. We detected objects with ranges up to $977$\,km and as small as $0.03 m^2$ radar cross section.  We found that 30 objects were either non-operational satellites or upper-stage rocket body debris. Additionally, we also detected FM reflections from Geminid meteors and aircraft flying over the MWA. Most of the detections of objects in LEO were found to lie within the parameter space predicted by previous feasibility studies, verifying the performance of the MWA for this application.  
\end{abstract}

\end{frontmatter}

\section{INTRODUCTION }
\label{sec:introduction}
With the advent of satellite mega-constellations, the density of objects in Low Earth Orbit (LEO) is predicted to reach $0.005 - 0.01$ objects per degree square \citep{mcdowell2020low}. Most of the current space surveillance radar systems dedicated to monitoring such objects in space (Space Domain Awareness: SDA\footnote{Previously Space Situational Awareness (SSA)}) operate at VHF/UHF/S-Band  and utilise active transmitters to reflect signals from objects in the space environment \citep{GOLDSTEIN19981007}. The predicted increase in the density of LEO objects demands detection systems with large instantaneous Field-of-View (FOV) receivers, the ability to change pointing directions and tracking quickly, and wide field illuminators. We aim to address these issues by using the Murchison Widefield Array (MWA) as a sensitive passive receiver in the FM band, coupled with existing, uncoordinated FM transmitters as the illuminators. 

Previously, \citet{prabu_hancock_zhang_tingay_2020} developed a technique using the so-called Dynamic Signal to Noise Ratio Spectrum (DSNRS) technique, that detects signals from satellites/debris, either via FM reflection or down-link transmission, and differentiates them from other types of Radio Frequency Interference (RFI) entering the detection system (the MWA). This previous work utilized the results of \citet{Zhang2018LimitsMWA} to select a set of MWA observations known to contain signals reflected from satellites. 

Having verified the DSNRS technique, we now take the next step in demonstrating SDA capabilities using the MWA, by developing and testing a semi-automated pipeline to perform uncued searches for the signals of interest. We test the pipeline on archival MWA data from observations in the FM band and we present here the results from the first low frequency non-coherent passive radar survey of LEO with the MWA. 
In this paper, we briefly summarise previous work in Section \ref{sec:background}. We describe our data processing pipeline in Section \ref{sec:dataReduction}, and our results in Section \ref{sec:results}. The discussion and conclusions are in Sections \ref{sec:discussion} and \ref{sec:conclusion}, respectively.

\section{BACKGROUND}
\label{sec:background}
Recently, many studies have raised concerns about the impacts of rapidly increasing LEO objects on astronomy \citep{mcdowell2020low,gallozzi2020concerns,hainaut2020impact,mallama2020flat}. We utilise this as an opportunity to demonstrate  space surveillance capabilities using an existing radio interferometer and terrestrial FM transmitters.

The MWA is a low frequency radio interferometer built as a precursor to the Square Kilometre Array (SKA) \citep{Tingay2013TheFrequencies}. The MWA can observe the sky at $70 - 300$\, MHz and was primarily designed for radio astronomy purposes \citep{2013PASA...30...31B, 2019arXiv191002895B}. The MWA has detected satellites in the past using two different techniques; namely coherent detection \citep{7944483,8835821} and non-coherent detection \citep{Tingay2013OnFeasibility,Zhang2018LimitsMWA,prabu_hancock_zhang_tingay_2020} methods. 

The coherent detection method uses the MWA's high time and frequency resolution Voltage Capture System (VCS) \citep{2015PASA...32....5T} and performs detections using matched filters designed using the transmitted FM signal \citep{8835821}, while the non-coherent detection system uses interferometer correlated data \citep{prabu_hancock_zhang_tingay_2020} along with wide-field imaging techniques. The blind detection pipeline developed here uses the non-coherent detection method, including the use of the DSNRS techniques established by \citet{prabu_hancock_zhang_tingay_2020}.

Electromagnetic simulations presented in \cite{Tingay2013OnFeasibility} predict that LEO objects with a radar cross section (RCS) greater than $0.79$\,$m^2$ and with line of sight (LOS) range less than $1000$\,km can be detected using the MWA in the FM band using non-coherent techniques, and we compare our obtained results with these predictions in Section \ref{sec:discussion}. 

\section{Data Processing}
\label{sec:dataReduction}
In this work, we aimed to autonomously search for signals from satellites in the MWA data using non-coherent techniques. We utilised observations that observed the sky in the frequency range $72.335-103.015$\,MHz, as this band partially overlapped with FM frequencies and a large number of observations in this band were readily available in the MWA archive. The $628$ observations (Table \ref{tab1}) used in this work were zenith pointing drift scans from four different nights performed using the MWA's phase 2 compact configuration \citep{2018PASA...35...33W}. The compact configuration has most of its baselines shorter than $200$\,m, thus enabling the detection system to be sensitive towards near-field objects at FM frequencies.



The visibility files for these observations were downloaded from the All-Sky Virtual Observatory \footnote{\url{https://asvo.mwatelescope.org/dashboard}} (ASVO) node for the MWA. They were converted to measurement sets  \citep{2007ASPC..376..127M} using COTTER \citep{2015PASA...32....8O} with a time averaging of $2$\,s and a frequency resolution of $40$\,kHz with RFI excision disabled.

Calibration observations were obtained as measurement sets from ASVO  and were preprocessed with AOFLAGGER \citep{2015PASA...32....8O}  to flag all baselines with RFI. This was followed by calibration of the measurement sets using the calibrator model. Once calibrated, in order to obtain calibration solutions for channels with RFI, we interpolate solutions between neighbouring channels.

After applying the interpolated calibration solutions to the target observations, the measurement sets were imaged at every time-step and fine frequency channel using WSCLEAN \citep{offringa-wsclean-2014,offringa-wsclean-2017}. WSCLEAN is the  abbreviation for W-Stack CLEANing, an advanced de-convolution method developed for wide-field interferometers. CLEAN (de-convolution) is usually done in order to reduce the side-lobes of the synthesised beam. However, we do not perform CLEAN as the subsequent step in our pipeline was to generate difference images, which remove the static celestial sources along with their side-lobes, revealing signals from objects such as satellites, meteors, and aircraft. 

\begin{table*}[h!]

\centering
\begin{tabular*}{\textwidth}{@{} c| c| c| c | c | l  @{}}
\hline \hline
Observation & Start  & End & Total Duration &  Calibration & Calibrator\\
IDs & UTC & UTC  & (hours) & Observation & Source  \\

\hline \hline

\hline \hline
 1157366872 - 1157407072 & 2016-09-08 10:47:34 & 2016-09-08 21:57:34 & 1.93 & 1157381872 & 3C444 \\
 \hline
 1157453032 - 1157493232 & 2016-09-09 10:43:34 & 2016-09-09 21:53:34 & 1.87 & 1157452432 & Her A \\
 \hline
 1160477632 - 1160507152 & 2016-10-14 10:53:34 & 2016-10-14 19:05:34 & 7.34 & 1160507272 & Pic A \\
 \hline
 1165749976 - 1165782976 & 2016-12-14 11:25:58 & 2016-12-14 20:35:58 & 8.4 & 1165779136 & Hyd A \\

\hline\hline
\end{tabular*}
\caption{List of observations and calibrator observations used in this work. Observation IDs can be searched within the MWA ASVO.}
\label{tab1}
\end{table*}

\subsection{BLIND SEARCH}
\label{sec:blindDetection}
\begin{figure*}[h!]
\begin{center}
\includegraphics[width=\linewidth,keepaspectratio]{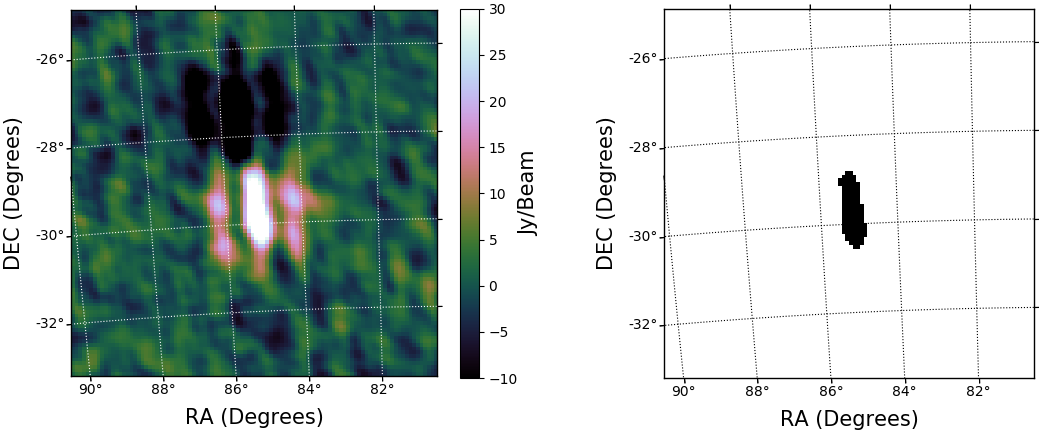}
\caption{The left panel shows a primary beam corrected $40$\,kHz fine channel difference image of KANOPUS-V. KANOPUS-V is an Earth observation mini satellite orbiting at an altitude of $510$\,km. The image shows two adjacent streaks caused by side-lobes. The right panel shows the floodfill region of the detected signal.}
\label{Fig1DiffImg}
\end{center}
\end{figure*}

After the images at every time-step and frequency channels were generated, a blind detection pipeline was run. The pipeline constructed difference images by subtracting the image at time-step $t$ from time-step $t+1$, for every fine frequency channel, and searched for pixels over $6\sigma$. The $6\sigma$ pixels were used to seed a detection, and we use a flood-fill\footnote{An algorithm that finds all adjacent cells that satisfy a given condition. We utilised the "forest fire algorithm" \citep{torbert2016applied}.} function to identify all adjacent pixels above $3\sigma$. An example of a satellite detected using this method is shown in Figure \ref{Fig1DiffImg}. The pixels together constitute the detected signal. We limit our algorithm to the detection of one event per time step per frequency, as when strong signals are present they are accompanied by many strong side-lobes, which we do not want to record as detections. Note that multiple detections at a single time-step are possible if they are seen in different frequency channels. Information for each detection, such as its coordinates (Right Ascension and Declination), peak flux density, time stamp, and frequency were stored for later analysis.

\subsection{Detection Maps}
\label{sec:detectionMaps}
\begin{figure*}[h!]
\begin{center}
\includegraphics[width=0.8\linewidth,keepaspectratio]{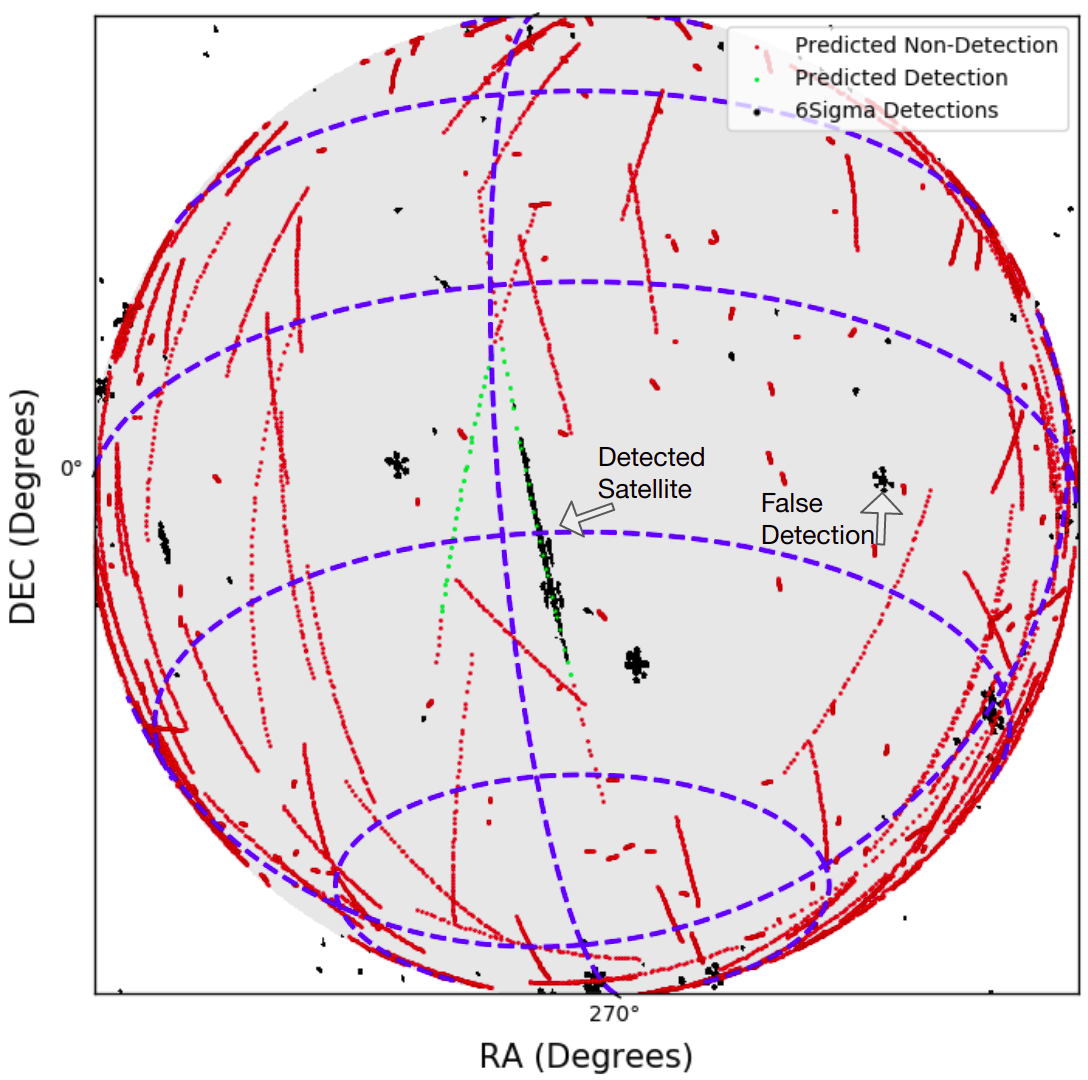}
\caption{The image shows the visible horizon during one of the $112$\,s MWA observations. The black markers are detections during this observation. The predicted orbits of all satellites within the visible horizon are plotted in red (or green). If the satellite orbit satisfies all predicted detection criteria (as predicted by \citet{Tingay2013OnFeasibility}) and is within MWA's half power beam, then its trajectory is plotted in green. One of the theoretically detectable satellites being detected by the pipeline is shown and one is not detected. There are several transmitters also detected near the horizon. The figure also shows one of the false detections that takes the shape of the point spread function.}
\label{Fig2detectionMap}
\end{center}
\end{figure*}

For each of the target observations, the positions of the detections were combined to make detection maps as shown in Figure \ref{Fig2detectionMap}. These detection maps are a visualisation tool to perform matching (by eye) of the detections in the observation with the predicted orbits of satellites in the FOV. In Figure \ref{Fig2detectionMap} the detections are shown in black. The predicted trajectories \footnote{Using TLE obtained from \url{https://www.space-track.org}} for all the objects in LEO, Middle Earth Orbit (MEO), and Higly Elliptical Orbits (HEO) above the horizon are plotted in red and green. \cite{Tingay2013OnFeasibility} predicts that the objects with range less than $1000$\,km and an RCS greater than $0.785$\, $m^2$ can be detected by the MWA. Hence, if the object is within the MWA's half power beam and satisfies the above mentioned conditions, then the red trajectory is replaced by green (as these are theoretically detectable orbits). The detections that were seen in multiple frequencies (in order to reduce the false positive events as described in Section \ref{sec:falsdetection}) can be  classified as satellites, meteor candidates, aircraft, terrestrial transmitters, unknown objects, and false detections, and are discussed in Section \ref{sec:results}.

\subsection{Parallax Analysis}
\label{sec:parallaxAnalysis}
The detections classified as aircraft (Section \ref{sec:planes}) appeared bright enough to be detected outside the MWA's primary beam and we estimate the range to these aircraft by performing parallax measurements. The MWA has 128 tiles, and splitting the array into two sub-arrays enables us to perform parallax measurements to some of these bright nearby events that are within the atmosphere.

The MWA compact configuration baselines were sorted in longitude, using the geometric centres of the baselines. Using this sorted list of baselines, the 1000 east-most baselines were combined to make an eastern aperture (ensemble of points in the UV plane) and the 1000 west-most baselines were combined to make a western aperture. The measurement sets for the eastern and western apertures were created by using the split\footnote{\url{https://casa.nrao.edu/docs/TaskRef/split-task.html}} task in Common Astronomy Software Applications (CASA\footnote{\url{https://casa.nrao.edu/}}) by providing the baseline configuration for both the apertures. 

Difference images for the full MWA compact array, eastern aperture, and western aperture were produced for one of the time-steps in which an aircraft was present. However, the UV coverages of the three apertures are different, resulting in different beams sizes. Hence, we address the problem by performing CLEAN and using a low resolution restoring beam corresponding to the lowest resolution of the three apertures. Due to the reflection signal being present in many frequency channels, we enabled the multi-frequency synthesis feature of WSCLEAN while imaging. The centres of the eastern and western apertures were calculated using the geocentric coordinates of the tiles obtained from the measurement set using casa-core \footnote{\url{https://casacore.github.io/python-casacore/}}. The two apertures result in a parallax baseline of $228.2$\,m.

\begin{figure}[h]
\begin{center}
\includegraphics[width=\linewidth,keepaspectratio]{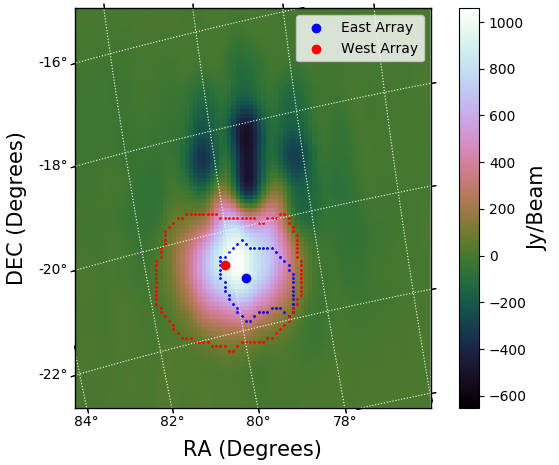}
\caption{$30.72$\,MHz bandwidth difference image of an aircraft using the MWA compact array. The blue and the red dotted lines are $3 \sigma$ contours of the streak when seen by the eastern and western apertures, respectively. The dots are the corresponding points of maximum brightness. Note that the contour of the eastern aperture image is smaller than that for the western aperture, due to the two sub-arrays having different sensitivities (number of short baselines) towards the aircraft's altitude. }
\label{Fig5parallax}
\end{center}
\end{figure}

The difference images made using the eastern and western apertures showed the parallax shift in the apparent position of the aircraft, as shown in Figure \ref{Fig5parallax}. Using the maximum brightness points and the centres of the two apertures, the LOS range to the aircraft was calculated as in \citet{earl2015determining} to be $20 \pm 2$\,km. The aircraft was detected at an azimuth of $82.6^\circ$ and an elevation of $26.3^\circ$, placing it at an altitude of $9 \pm 1$\,km (height of most civil aircraft). Note that although the baselines were sorted in longitude to maximise the East-West separation, the centres of the two apertures have a latitude component as well, thus in Figure \ref{Fig5parallax} we see a combination of East-West and North-South offsets in the apparent position.

\section{Results}
\label{sec:results}
\subsection{Satellite Candidates}
\label{sec:satellites}
\begin{figure}[h!]
\begin{center}
\includegraphics[width=\linewidth,keepaspectratio]{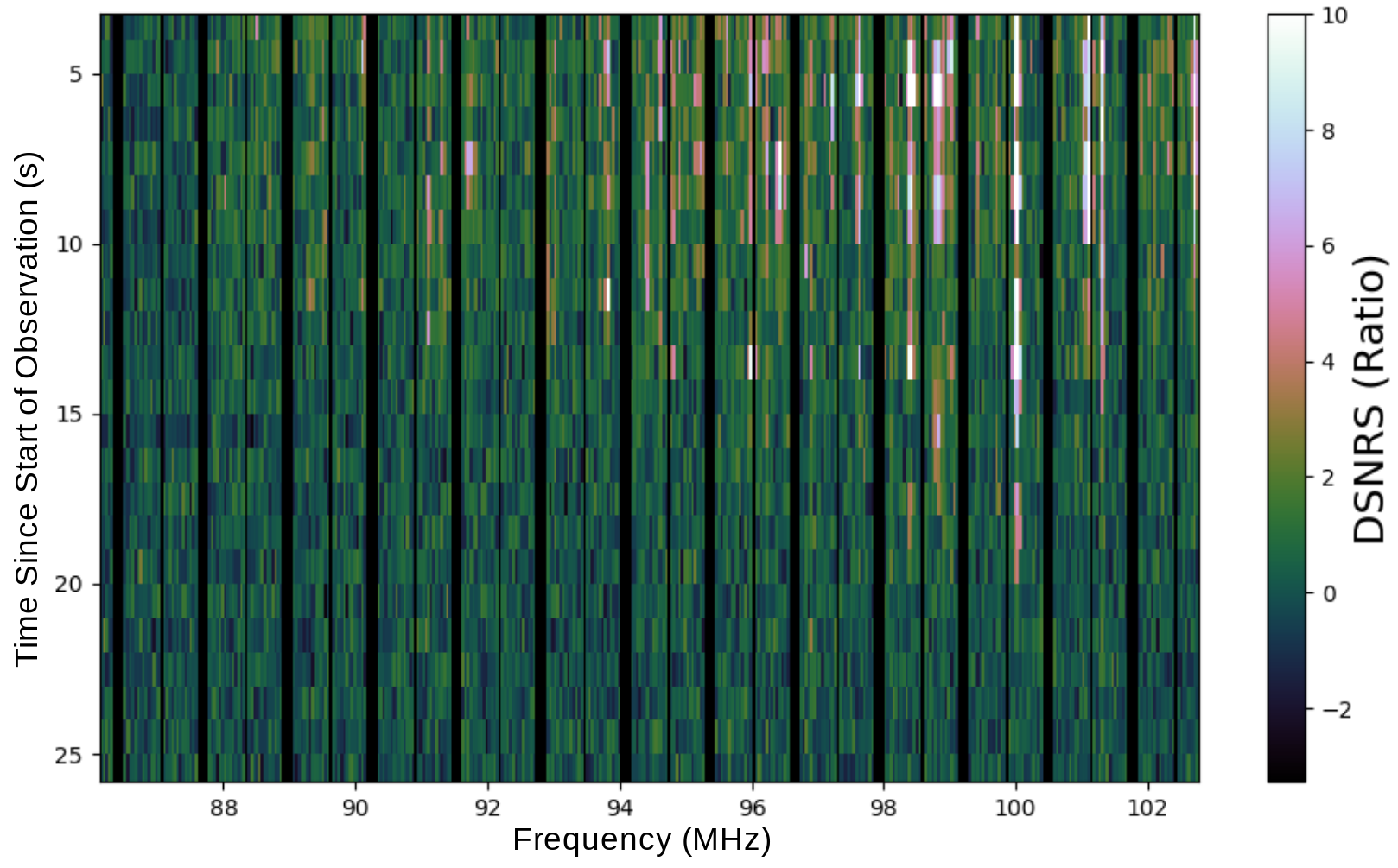}
\caption{DSNRS plot for ZIYUAN 3 (ZY 3). The plot shows the different FM frequencies reflected by the satellite. The black vertical lines in the figure are due to the flagging of trailing, central, and leading fine frequency channels in every coarse channel.}
\label{Fig3dsnrs}
\end{center}
\end{figure}

Visual inspection of the detection maps for each of the observations was performed, and the events that plausibly matched in time and position with known objects at multiple time-steps were classified as satellite candidate detections. A total of 74 unique LEO objects were detected over multiple passes, of which 15 were upper stage rocket body debris. The LOS ranges for these satellites were obtained for the time-steps they were detected (calculated using the Two Line Element (TLE) values). The range values, along with RCS, peak flux densities, and operational statuses for these detected objects are tabulated in Table \ref{tab2}.  An example DSNRS plot, illustrating the range of frequencies and times for which a satellite was detected is shown in Figure \ref{Fig3dsnrs}.

Two satellites, the CubeSats DUCHIFAT-1 and UKUBE-1, were detected due to out-of band transmissions in the FM band, rather than reflections (as previously observed by \cite{Zhang2018LimitsMWA} and  \cite{prabu_hancock_zhang_tingay_2020}. 

\begin{table*}[h!]
\caption{Detected Satellites/Debris and their properties.}
\label{tab2}
\centering
\begin{tabular*}{\textwidth}{@{} c\x c\x c\x c\x c\x c\x c\x c\x@{}}
\hline \hline
Observation & NORAD & Satellite/Debris  & Range & RCS & Operational& $\theta$ & Peak Flux\\
ID & ID & Name & $km$ & $m^2$ & Status  & Degrees &  Density (Jy/beam)\\
\hline 
\hline
\multicolumn{8}{c}{The detections below are from the night of 2016-12-14 from 11:25:58 UTC to 20:35:58 UTC}\\
\hline \hline
1165782616 & 33408  & SJ-6E& 598 - 603 & 1.3 &O&    5.0&32.5\\
\hline
1165782016 & 28898 & MOZHAYETS 5 and RUBIN-5 & 699 - 709 & 5.9 &N/A& 2.3&31.2\\
\hline
1165780696 & 23088 & SL-16 R/B  & 863 - 873 & 10.3 & R/B & 13.4 & 137.9\\ 
\hline
1165779376 & 13367 &LANDSAT 4 & 538 - 539 &  6.4 &NO & 6.3 &35.3\\
\hline
1165777336 & 28230 & GP-B & 669 - 687 &  10.2 & NO& 9.9 &61.3\\
\hline
1165777216 & 9786 & DELTA 1 R/B(1) & 621 - 624 & 8.9 & R/B &6.5 &55.4 \\
\hline
1165776496 & 40420 &COSMOS 2503 & 587 - 600 &  5.5 &O & 3.1&33.6\\
\hline
1165773496 & 40310 & YAOGAN 24& 639 - 656  & 4.2 &O & 3.3 &115.8\\
\hline
1165773136 & 24277 & MIDORI (ADEOS) & 806 - 816 & 22.2 &NO & 5.8 &51.2\\
\hline
1165772296 & 13153 & COSMOS 1356 & 480 - 486 & 9.0 & N/A& 7.3 &39.9\\
\hline
1165771216 & 33492 & GOSAT (IBUKI) & 681 - 705  & 4.6 &O & 5.1 &56.3\\
\hline
1165771096 & 33053 &FGRST (GLAST) & 555 - 563 & 4.9 &E& 17.7&233.9\\
\hline
1165770136 & 41336 & BREEZE-KM R/B & 534 - 540 &3.3 & R/B & 20.0 & 38.2\\
\hline
1165768696 & 20580 &HST& 617 - 669 &28.1&O&25.2&449.5\\
\hline
1165767856 & 25078 &IRIDIUM 44& 780 - 783 & 3.3 &NO& 6.2&44.5\\
\hline
1165766176 & 38707 &KANOPUS-V 1& 518 - 543 & 1.9 &O&5.7 &81.9\\
\hline
1165765696 & 41731 &QSS (MOZI)& 528 -  540 & 2.1 &O& 20.8 &50.5\\
\hline
1165765336 & 39152 &TURKSAT-3USAT & 632 - 642 & 0.1 &NO&3.9 &25.5\\
\hline
1165765216 & 25544 &ISS (ZARYA)& 651 - 877 & 399.1 &O& 47.2 & 247,009\\
\hline
1165765096 & 25544 &ISS (ZARYA)& 733 - 977  & 399.1 &O& 44.4 & 	25,936\\
\hline
1165764136 & 25758 &IRS-P4 (OCEANSAT) & 730 - 764 & 3.5 &NO& 0.9&44.8 \\
\hline
1165764136 & 28499 & ARIANE 5 R/B & 687 - 698 &16.0 & R/B &13.0  &45.1\\
\hline
1165763056 & 39019 &PLEIADES 1B& 719 - 729 & 5.4 &O& 8.2&26.7\\
\hline
1165762576 & 20580 &HST& 578 - 600  & 28.1 &O& 19.9&120.6\\
\hline
1165761856 & 41848 &WORLDVIEW-4& 626 - 634 & 6.6 &PO&5.3&30.2\\
\hline
1165761736 & 27601 &H-2A R/B &844 - 879 &24.6 & R/B & 5.9& 34.8\\
\hline
1165761376 & 41341 &H-2A R/B & 576 - 607 &27.4 & R/B & 8.5 & 127.1\\
\hline
1165761256 & 38046 &ZIYUAN 3 (ZY 3)& 528 - 561 & 5.3 &O& 13.7&139.3\\
\hline
1165761136 & 38046 &ZIYUAN 3 (ZY 3)& 513 - 562 & 5.3 &O&5.8 &152.1\\
\hline
1165760896 & 21422 &COSMOS 2151& 618 - 625 & 5.7 &N/A&9.2&26.5\\
\hline
1165760776 & 12987 &COSMOS 1328& 565 - 579 &8.2&N/A&6.5&44.2\\
\hline
1165760536 & 38249 &PSLV R/B & 381 - 407 &5.8 & R/B &1.4 &55.0\\
\hline
1165758616 & 29499 &METOP-A& 862 - 878 & 11.2 &O& 17.1&53.0\\
\hline
1165757056 & 27386 &ENVISAT& 782 - 805 & 18.6 &NO& 8.0&117.6\\
\hline
1165756576 & 20580 &HST& 565 - 584  & 28.1 &O& 13.4&59.0\\
\hline
1165756576 & 29228 &RESURS-DK 1& 583 - 596 & 8.8 &O&13.0&25.7\\
\hline
1165756456 & 20580 &HST& 551 - 553 & 28.1&O&8.3&80.8\\
\hline
1165756096 & 11060 &TIROS N& 849 - 853 & 4.1&PO& 0.8&37.2\\
\hline
1165755976 & 14819 &COSMOS 1544& 505 - 526  & 8.3&N/A&0.3&179.8\\
\hline
1165754896 & 32062 &CBERS 2B& 773 - 784  & 2.5 &NO& 12.7&38.4\\
\hline
1165753936 & 16881 &COSMOS 1766& 558 - 584  & 8.3 &N/A& 3.3& 52.3\\
\hline
1165753936 & 23968 &ATLAS 2 CENTAUR R/B & 472 - 528 &14.9& R/B & 5.2&242.6 \\
\hline
1165752856 & 16613 &SPOT 1& 691 - 702  & 7.3&NO& 15.4 &102.4\\
\hline \hline
\multicolumn{8}{c}{The detections below are from the night of 2016-10-14 from 10:53:34 UTC to 19:05:34 UTC}\\
\hline \hline
1160505472 & 38257 & YAOGAN 14  & 493 - 505 & 5.41 & O& 7.5 & 143.0 \\
\hline
1160504512 & 10490 & DELTA 1 R/B(1)& 523 - 530 & 9.1 & R/B &9.3 &48.1  \\
\hline

\end{tabular*}\label{blah}
\tabnote{Continued on next page...}
\end{table*}

\addtocounter{table}{-1}
\begin{table*}[h!]
\caption{...continued from previous page.}
\centering
\begin{tabular*}{\textwidth}{@{} c\x c\x c\x c\x c\x c\x c\x c\x@{}}
\hline \hline
Observation & NORAD & Satellite/Debris  & Range & RCS & Operational& $\theta$ & Peak Flux\\
ID & ID & Name & $km$ & $m^2$ & Status  & Degrees &  Density (Jy/beam)\\
\hline 
\hline
1160504752 &24796 &IRIDIUM 4  & 805 - 815 & 3.7 & NO & 13.5 &63.1 \\
\hline
1160502952 & 21574 &ERS-1  & 790 - 794 & 10.3 & NO & 4.4 & 32.7\\
\hline
1160502472 & 15427 & NOAA 9 & 876 - 904  & 4.3 & PO & 13.4 &61.8 \\
\hline
1160500432 & 28480 & CZ-2C  & 841 - 843 & 10.0 & R/B & 13.4 &38.2\\
\hline
1160498872 & 36095 & COSMOS 2455 & 914 - 917 & 12.2 & O &2.9 &41.8\\
\hline
1160497792 & 24950 & IRIDIUM 31 & 793 - 800  & 3.6 & N/A & 2.3 &35.1 \\
\hline
1160497672 & 25544 & ISS (ZARYA) & 454 - 577 & 399.1 &O &21.5 &23,492\\
\hline
1160497672 & 40074 & UKUBE-1 & 739 - 760 & 0.1 & O &31.3 & 417.1 \\
\hline
1160497552 & 25544 & ISS (ZARYA) & 442 - 587 & 399.1 & O&17.9 &19,138 \\
\hline
1160497192 & 19274 & OKEAN-1 & 573 - 586 &   8.6 & N/A &7.7 & 49.6\\
\hline
1160497072 & 19274 & OKEAN-1 & 564 - 575 & 8.6 & N/A &4.0 &  40.2\\
\hline
1160497072 & 41386 & RESURS P3 & 520 - 546   & 7.7 & O&18.5 &160.2 \\
\hline
1160496352 & 39574 & GPM-CORE & 415 - 435 & 8.1 & O &11.1 &70.2 \\
\hline
1160496232 & 39574 & GPM-CORE & 410 - 479  & 8.1 & O& 7.0&  598.1 \\
\hline
1160495752 & 23608 &  ARIANE 40+3 R & 602 - 619 & 9.7 & R/B &6.8 & 142.6\\
\hline
1160493592 & 40118 & GAOFEN 2 & 642 - 714 & 3.5 & O & 3.4 & 126.3\\
\hline

1160493472 & 40021 & DUCHIFAT-1  & 647 - 709 & 0.03 & O &17.6 & 469.0\\
\hline
1160493472 & 25260 & SPOT 4 & 716 - 752 & 6.2 & NO &0.5 & 109.4\\
\hline
1160492512 & 28649 & IRS-P5 (CARTOSAT-1) & 654 - 668 & 4.7 & O & 15.4 & 105.6\\
\hline
1160492392 & 28649 & IRS-P5 (CARTOSAT-1) & 640 - 647 & 4.7 & O & 10.2 & 44.9\\
\hline
1160491192 & 20624 & COSMOS 2082 & 864 - 888 & 10.8 & N/A & 11.1 &  146.4\\
\hline
1160490232 & 23697 & ATLAS 2 CENTAUR& 919 - 929 & 13.9 & R/B &2.5 & 76.0\\
\hline
1160489512 & 812 & OPS 4467 A & 821 - 844 & 0.34 & N/A & 0.9 &48.9 \\
\hline
1160488792 & 27421 & SPOT 5 & 659 - 665 & 7.3 & NO &  9.4 & 130.6\\
\hline
1160487952 & 41765 & TIANGONG-2 & 446 - 455 & 15.8 & N/A & 28.4 & 232.9 \\
\hline
1160487832 & 23317 & OKEAN-4 & 639 - 656 & 7.1 & N/A & 9.5 & 208.3 \\
\hline
1160486632 & 8845 & METEOR 1-25& 884 - 896 & 4.0 & N/A &8.6 & 122.7 \\
\hline
1160485792 & 39358 & SHIJIAN-16 (SJ-16) & 643 - 646 & 8.3 & O &  12.9 & 147.9 \\
\hline
1160484112 & 28118 &ATLAS 3B CENTAUR & 313 - 349 & 11.9 & R/B & 10.3 &174.6\\
\hline
1160479192 & 40913 &CZ-6 R/B & 460 - 465 &2.6 &R/B &7.5 & 61.4\\
\hline \hline
\multicolumn{8}{c}{The detections below are from the night of 2016-09-09 from 10:43:34 UTC to 21:53:34 UTC} \\
\hline\hline
1157493232 & 41727& GAOFEN 3 & 790 - 811  &3.9 & O & 14.7 &256.4 \\
\hline
1157486032 & 19549 &IUS R/B(1) &  298 - 303 & 11.8 & R/B & 14.9 & 1606 \\
\hline
1157474632 & 20580 &HST & 551 - 583 & 28.1 & O  &9.3 & 1336.2 \\
\hline

1157472832 & 35931 & OCEANSAT-2  & 731 - 741 & 4.1 & O &2.9 & 113.7\\
\hline
1157472832 & 41386 & RESURS P3 & 479 - 489  & 7.7 & O &1.6 & 121.9 \\
\hline
1157468632 & 20580 & HST  & 590 - 633 & 28.1 & O &22.0 & 1306 \\
\hline\hline
\multicolumn{8}{c}{The detections below are from the night of 2016-09-08 from 10:47:34 UTC to 21:57:34 UTC}\\
\hline\hline
1157407072 &  41456 & SENTINEL-1B& 738 - 754  & 5.6 &O & 15.7 & 77.7\\
\hline
1157407072 & 32382 & RADARSAT-2 &  804 - 812 & 8.4 & O & 4.4 & 45.0 \\
\hline
1157394472 & 41026 & YAOGAN 28 & 505 - 563 & 4.8 & O &16.5 &672.3  \\
\hline
1157393872 & 20978& DMSP 5D-2 F10 (USA 68)&  840 - 846 & 3.9 & NO & 16.4 &47.4 \\
\hline
1157383672 & 33504 & KORONAS-FOTON & 545 - 547 & 4.2 &NO &2.6 & 47.2 \\
\hline
1157382472 & 15944&  COSMOS 1674& 546 - 570 & 8.7 & N/A & 10.2 & 44.3 \\
\hline
\hline
\end{tabular*}\label{blah}
\tabnote{Legend: O=Operational, R/B=Rocket Body, NO=Non-Operational, PO=Partially Operational, N/A=Not Available.
The table summarises the properties of all the detected satellites. It provides the satellite’s North American Aerospace
Defence (NORAD) ID, the range of distance over which it was detected, its Radar Cross Section (RCS\footnote{Obtained from \url{https://celestrak.com/pub/satcat.txt}}), the zenith angle $(\theta)$, and the primary
beam corrected peak flux density as seen in the brightest 40 kHz frequency channel. Note that the operational status \footnote{\url{https://celestrak.com/satcat/satcat-format.php}}
may not be accurate as the information source does not list the date it was last updated. Note that the Observation ID is the GPS time of the start of the observation.}
\end{table*}

\subsection{Meteor Candidates}
\label{sec:meteors}
\begin{figure}[h!]
\begin{center}
\includegraphics[width=\linewidth,keepaspectratio]{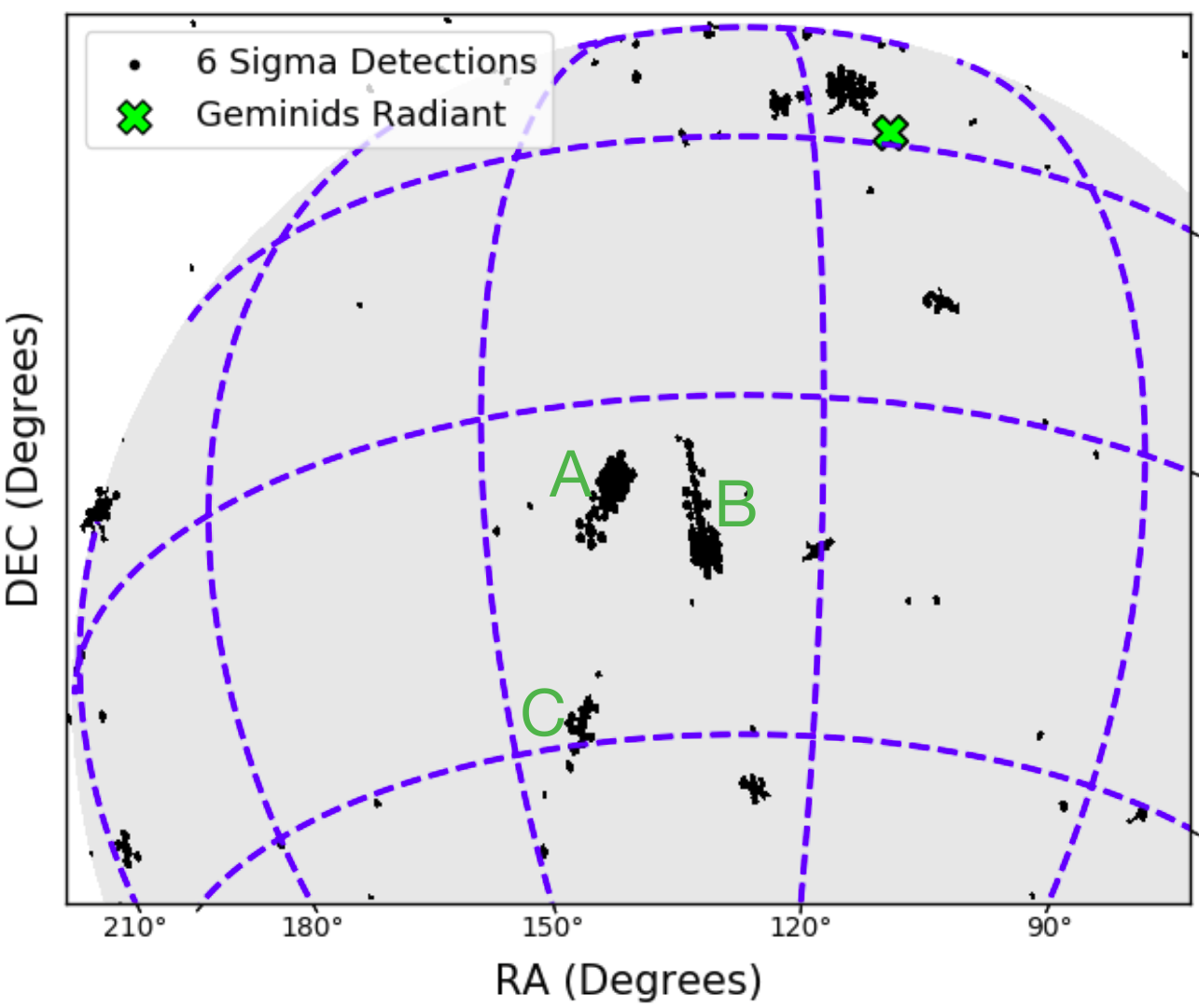}
\caption{Three of the detected meteors are shown in regions A, B and C. Meteor-A and meteor-C point in the direction of the Geminids Radiant while meteor-B could be a sporadic meteor. }
\label{Fig4meteor}
\end{center}
\end{figure}

The observations from one of the nights used in this work (14th December 2016) coincided with the  Geminids meteor shower. The pipeline detected many reflections from objects that had angular speeds much greater than expected for LEO objects. These objects moved approximately $10$ degrees in a single $2$\,s time-step and are FM reflections from the ionised trails of meteors, as previously observed by \citet{Zhang2018LimitsMWA} with the MWA. An example is shown in Figure \ref{Fig4meteor}. These events often appeared much brighter than satellites and were often pointing in the direction of the Geminids radiant.

\subsection{Aircraft}
\label{sec:planes}
 Nineteen aircraft passes were detected by the pipeline, due to their large reflecting areas and smaller ranges. Most of these aircraft flew North-South over the MWA (a very common flight path for flights between Singapore/Malaysia/northern WA locations and Perth). These reflections appeared very bright (approximately $2800$\,Jy/beam peak flux density in a $30.72$\,MHz bandwidth difference image) and we utilised parallax to determine their altitudes (Section \ref{sec:parallaxAnalysis}).

\subsection{Transmitters and Unknown Objects}
\label{sec:transmittersAndUnkownObjects}
Transmitters near the horizon were often detected. These transmitters are not removed through difference images as they are at a fixed azimuth and elevation, hence appear to move in celestial coordinates with time. In future observations, these azimuths/elevations will be masked in order to prevent the pipeline from detecting these transmitters. The transmitters are seen at multiple FM frequencies.

We also detected several events that had angular speeds very similar to LEO objects but did not coincide with any known orbits in the TLE catalog. These are likely to be either satellites with outdated TLEs or uncatalogued objects (intentionally or otherwise). In future, we will investigate these events further by performing orbit determination estimates.

\subsection{False Positives}
\label{sec:falsdetection}
The noise in difference images mainly consists of thermal noise and is assumed to follow Gaussian statistics. Due to the large volume of data used in this work, thermal noise fluctuations can trigger the $6 \sigma$ threshold of the detection pipeline, and hence it is important that we quantify these false positives. However, since we constrain the pipeline to allow only the brightest detection per time-step and per frequency channel, the number of false detections is reduced in the presence of a bright reflection event that is seen in multiple frequencies.

In order to investigate the number of false positives, we ran our pipeline again but only on the 380 fine channels outside the FM band (i.e outside $87.5 - 108$\,MHz, which is the FM band in Australia). By doing so, we only detect the false positives as the reflection events are confined to the FM band. Note that observations that had no transmitting satellites were used for this analysis, as the transmitted signals from satellites were not confined to the FM band. 

We obtained an average of 13 false detections per minute, for the 380 fine frequency channels used. Thus for a full bandwidth observation, and in the absence of any satellite detection, we would obtain approximately 26 false detections per minute. However, since we utilise other tools such as DSNRS (frequency and time analysis) and detection maps (position and time analysis), to investigate these events further, the probability of classifying one of these events as a LEO object is insignificantly small.

\section{DISCUSSION}
\label{sec:discussion}
\subsection{Detection Completeness}
\label{sec:detectionCompleteness}

\begin{figure*}[h!]
\begin{center}
\includegraphics[width=\linewidth,keepaspectratio]{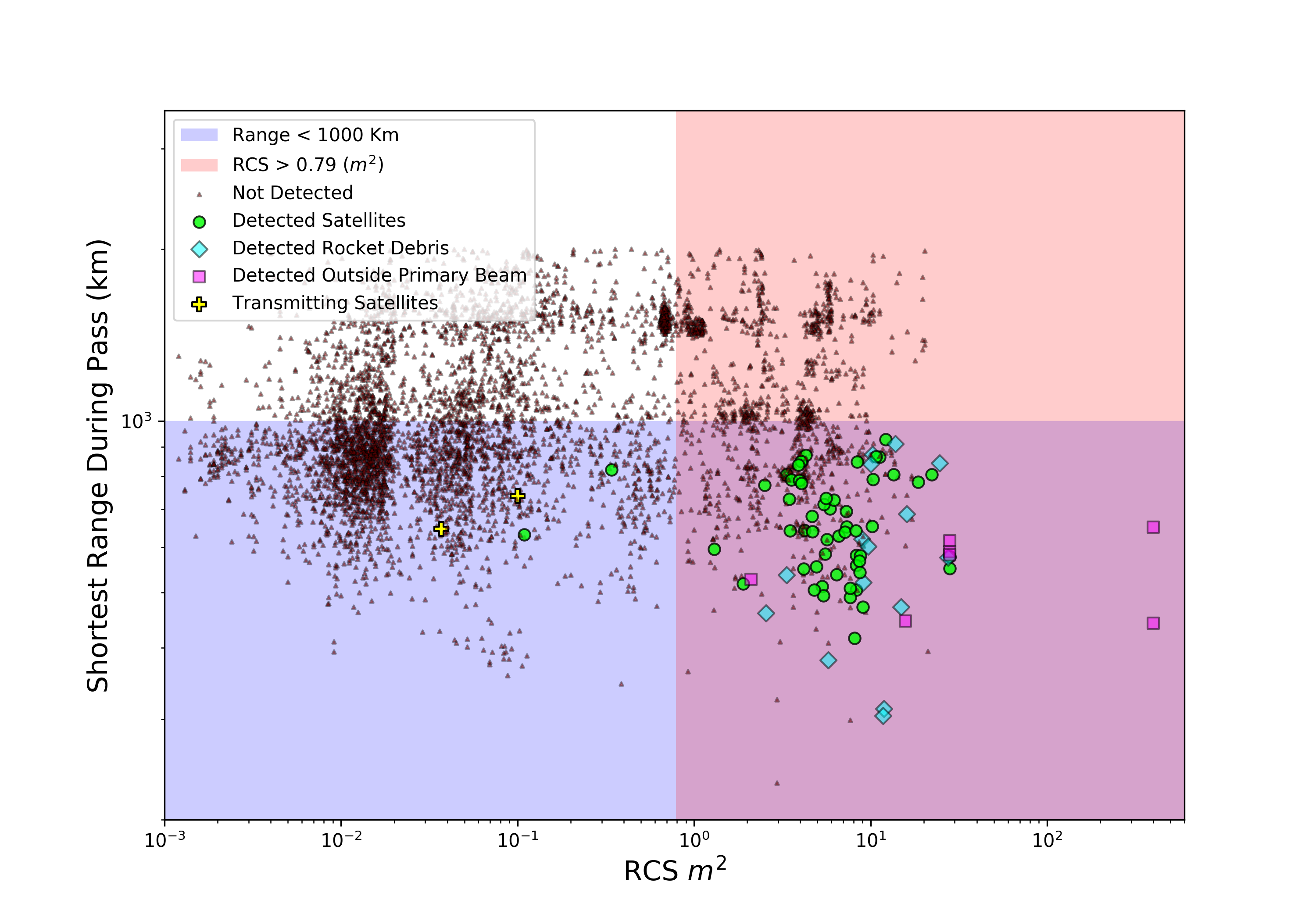}
\caption{The RCS and the shortest range for all the satellites/debris passes above the horizon within the half power beam and with a range less than $2000$\,km. Note that although a satellite can appear in two consecutive observation IDs, it appears in the above plot as a single datum e.g. the ISS is detected in four observations according to Table \ref{tab2}, but only appears twice in the above plot (two rightmost points with the largest RCS) because those four observations covered two passes.}
\label{Fig6detectionCompleteness}
\end{center}
\end{figure*}

\citet{Tingay2013OnFeasibility} predicts that satellites with an RCS greater than $0.79$\,$m^2$ and with LOS range less than $1000$\,km can be detected using the MWA in the FM band using non-coherent techniques. All the satellites/debris that passed through the MWA's half power beam with a shortest range during a pass less than $2000$\,km were identified and their RCS, along with the shortest range during pass, are plotted in Figure \ref{Fig6detectionCompleteness}. All of the detected objects in this work (except three CubeSats and one MiniSat) were detected within the theoretically predicted parameter space. Two of the CubeSats (DUCHIFAT-1 and UKUBE-1) were detected due to out-of band transmissions in the FM band (as previously observed by \cite{Zhang2018LimitsMWA} and \cite{prabu_hancock_zhang_tingay_2020})  and the other CubeSat and MiniSat were detected through FM reflections. Some satellites such as the ISS and Hubble Space Telescope (HST) were also detected outside the MWA's primary beam due to their large RCS.

From Figure \ref{Fig6detectionCompleteness} it can be seen that not all the satellites in the predicted parameter space where detected. This could be due to a number of reasons, for example unfavourable reflection geometries, or our pipeline being constrained to allow only one detection per time step per frequency channel. One significant reason could be that the RCS values are estimated by the US Space Surveillance Network (SSN)\citep{SSN} using VHF/UHF/S-Band radars and are very likely to be quite different at the FM frequencies considered in this work. The RCS can also vary drastically as the transmitter-target-MWA reflection geometry changes and as the satellite tumbles. Also, the radar measured RCS is usually for a direct back-scatter/reflection where the transmitter and the receiver are co-located, as opposed to our method where we are looking at an oblique scattering of radiation (bi-static radar).  Hence, we use the cataloged RCS values as an order of magnitude guide only. Also, since the classification of an event as a LEO object is done by visual inspection, it is possible that we missed  detections near the horizon as it is usually crowded with many orbits due to projection effects as seen in Figure \ref{Fig2detectionMap}. 

From Table \ref{tab2} we can see that many satellites, such as the HST, were detected multiple times on the same night, demonstrating the MWA's re-acquisition capability for large objects. Many objects such as rocket body debris and non-operational satellites were also detected, and for these objects passive space surveillance is the only way we can track them, thus demonstrating the MWA's utility to track large obsolete objects. One such example is the object OPS 4467 A (NORAD ID 812). This satellite is the oldest object detected in our work and was launched in 1964. 

Other interesting detected objects from Table \ref{tab2} are MOZHAYETS-5 and RUBIN-5, which were launched together on the same rocket.  RUBIN-5 was designed to stay attached to the payload adapter while MOZHAYETS-5 failed to detach from the adapter and hence they appear together as a single object in Table \ref{tab2}. 

In one of the observations, the ISS was detected near the horizon with a peak flux density of $247,009$\,Jy/beam in one of the $40$\,kHz fine frequency channels. This could be due to a favourable reflection geometry and reflections from its very large solar panel arrays.



\section{CONCLUSIONS}
\label{sec:conclusion}
We have built upon previous work using the MWA as a passive radar system by developing a semi-automated pipeline that searches for reflected signals from LEO satellites in high time and frequency resolution data. Previous detections were performed by manual inspection of full band-width difference images, and here we have dramatically increased the number of detections by searching autonomously in every fine frequency channel. 

Testing our pipeline on archived MWA data, we detected more than 70 unique LEO objects in 20 hours of observation. DUCHIFAT-1 and UKube-1 were detected due to spurious transmissions, while every other detected object was due to FM reflections. The large number of satellite detections through FM reflections alone prove MWA to be a valuable future asset for the global SDA network. 

All, except four, of the detected objects were found to lie within the parameter space (range vs RCS) predicted by \cite{Tingay2013OnFeasibility}. However, not all objects that were predicted to be detectable were detected. This could be due to a number of reasons such as tumbling and unfavourable reflection geometries reducing the RCS of the object. 

Along with the many satellite detections, we also detected FM reflections from Geminid meteors and aircraft flying over the MWA. Some detected events had angular speeds similar to LEO objects but did not have a satellite orbit match. In the future, we will further examine these unidentified objects by performing orbit determination. We will also use our data to demonstrate a detailed LEO catalog maintenance capability. The Gauss orbit determination technique \citep{curtis2013orbital} will be utilised, as we only measure the angular migration of the objects with non-coherent techniques. In future, the detection pipeline used here will be upgraded to preform fully autonomous detections instead of the visual inspection performed here. 

Many satellites transmit at MWA frequencies for down-link telemetry. Hence, observing in these frequencies could expand our detection window beyond the feasible parameter space (RCS-range) shown in this work.  The future detection and characterisation of satellites that unintentionally transmit out of band will also assist in determining the threat of mega-constellations of small satellites to ground-based radio astronomy facilities. 

\begin{acknowledgements}

This scientific work makes use of the Murchison Radio-astronomy Observatory, operated by CSIRO. We acknowledge the Wajarri Yamatji people as the traditional owners of the Observatory site. Support for the operation of the MWA is provided by the Australian Government (NCRIS), under a contract to Curtin University administered by Astronomy Australia Limited. We acknowledge the Pawsey Supercomputing Centre which is supported by the Western Australian and Australian Governments. Steve Prabu would like to thank Innovation Central Perth, a collaboration of Cisco, Curtin University,  Woodside and CSIRO’s Data61, for their
scholarship.

\subsection*{Sofware}
We acknowledge the work and the support of the developers of the following Python packages:
Astropy \citep{theastropycollaboration_astropy_2013,astropycollaboration_astropy_2018}, Numpy \citep{vanderwalt_numpy_2011}, Scipy  \citep{jones_scipy_2001}, matplotlib \citep{Hunter:2007} and Ephem\footnote{\url{https://pypi.org/project/ephem/}}. The work also used WSCLEAN \citep{offringa-wsclean-2014,offringa-wsclean-2017} for making fits images and DS9\footnote{\href{http://ds9.si.edu/site/Home.html}{ds9.si.edu/site/Home.html}} for visualization purposes. 

\end{acknowledgements}

\bibliographystyle{pasa-mnras}
\bibliography{custom}

\end{document}